\documentclass[letterpaper]{article} 
\usepackage{aaai2026}  
\usepackage{times}  
\usepackage{helvet}  
\usepackage{courier}  
\usepackage[hyphens]{url}  
\usepackage{graphicx} 
\usepackage{natbib}  
\usepackage{caption} 
\usepackage{algorithm}
\usepackage{algorithmic}

\usepackage{newfloat}
\usepackage{listings}
\DeclareCaptionStyle{ruled}{labelfont=normalfont,labelsep=colon,strut=off} 
\floatstyle{ruled}
\newfloat{listing}{tb}{lst}{}
\floatname{listing}{Listing}
\usepackage{amsmath}
\usepackage{multirow}
\usepackage{subcaption}
\usepackage{bm}

\title{Failing on Bias Mitigation: A Case Study on the Challenges of Fairness in Government Data}

\author {
    Hongbo Bo\textsuperscript{\rm 1},
    Jingyu Hu\textsuperscript{\rm 1},
    Debbie Watson\textsuperscript{\rm 2},
    Weiru Liu\textsuperscript{\rm 1}
}
\affiliations {
    \textsuperscript{\rm 1}School of Engineering Mathematics and Technology, University of Bristol\\
    \textsuperscript{\rm 2}School for Policy Studies, University of Bristol\\
    hongbo.bo@bristol.ac.uk,
jingyu.hu@bristol.ac.uk,
debbie.watson@bristol.ac.uk,
weiru.liu@bristol.ac.uk
}

\nocopyright
\begin{document}

\maketitle

\begin{abstract}
The potential for bias and unfairness in AI-supporting government services raises ethical and legal concerns. Using crime rate prediction with the Bristol City Council data as a case study, we examine how these issues persist. Rather than auditing real-world deployed systems, our goal is to understand why widely adopted bias mitigation techniques often fail when applied to government data. Our findings reveal that bias mitigation approaches applied to government data are not always effective
-- not because of flaws in model architecture or metric selection, but due to the inherent properties of the data itself. Through comparing a set of comprehensive models and fairness methods, our experiments consistently show that the mitigation efforts cannot overcome the embedded unfairness in the data -- further reinforcing that the origin of bias lies in the structure and history of government datasets. 
We then explore the reasons for the mitigation failures in predictive models on government data and highlight the potential sources of unfairness posed by data distribution shifts, the accumulation of historical bias, and delays in data release.
We also discover the limitations of the blind spots in fairness analysis and bias mitigation methods when only targeting a single sensitive feature through a set of intersectional fairness experiments.
Although this study is limited to one city, the findings are highly suggestive, which can contribute to an early warning that biases in government data may persist even with standard mitigation methods. 
\footnote{This manuscript is an extended version of the paper accepted at ICCART 2026, including additional experiments and appendices.}
\end{abstract}

\section{Introduction}

The recent UK Labour Government's Artificial Intelligence (AI) Opportunities Action Plan~\cite{UKGov2025AIActionPlan} makes ambitious promises about scaling AI applications across public services. The application of AI technology on government data has indeed attracted attention in recent years, as AI can help better provide public services~\cite{alshahrani2024artificial, alhosani2024opportunities}. For example, using AI models to predict crime based on government statistical data can help local governments optimize resource allocation and improve public security~\cite{safat2021empirical, dakalbab2022artificial}. However, the potential bias and unfairness in the AI decision-making process are particularly sensitive in government services and may even trigger ethical and legal issues~\cite{jaime2024ethnic, zezulka2024fair}, such as the public outcry concerning potentially unfair predictions in algorithms used for child protection~\cite{moreau2024failing, guardian2023thinkfamilyapp}. Critics point out that prioritizing AI expansion without sufficient regard for the underlying risks can lead to social inequalities and potentially exacerbate far-right ideologies~\cite{McQuillan2025LabourAI}. This makes the urgent to examine how the deployment of AI in government and public sector applications can affect social fairness.

The fairness of AI systems in government decision-making has been the focus of numerous studies, such as whether public resources like education or welfare are evenly distributed~\cite{barocas2023fairness}, whether algorithmic decision-making processes are fair~\cite{pessach2022review}, and whether the AI system provides fair decisions for all groups without bias against disadvantaged groups~\cite{barocas2023fairness, buolamwini2018gender}. In this study, we focus on exploring the unfairness against disadvantaged groups of different races and religions in AI systems for government decisions, through exploring a crime rate prediction study as a case to illustrate how the unfairness problem persists. We train crime rate predictive models based on a public dataset from Bristol City Council in South West England. Although the predictive models perform well, we find unfairness issues in the prediction results, the models show large error differences in results between the samples with high or low values of a sensitive feature, such as a proportion of race or religion. This result reveals that some machine learning models can be unfair when dealing with sensitive features from government data, and if the government's decisions are made based on unfair AI support tools, it will cause ethical issues, such as the overrepresentation of Black and minoritized in violence problems~\cite{pastaltzidis2022data}. 

Many research studies focus on mitigating bias and unfairness in machine learning models, which mainly include pre-processing methods, in-processing methods and post-processing methods. Pre-processing mitigations aim to provide a fair training dataset by balancing data samples~\cite{he2009learning, zhang2017mixup} and reducing sensitive features~\cite{feldman2015certifying}. In-processing methods focus on mitigating the unfairness during the model training process by adding fairness constraints~\cite{berk2017convex} or optimizing the loss function~\cite{bellamy2019ai} to reduce the impact of sensitive features. Post-processing mitigations are applied to correct the outputs~\cite{pleiss2017fairness} to ensure fairness after a model has been trained. In this study, the pre-processing and in-processing methods~\cite{he2009learning, zhang2017mixup, zafar2017fairness, hendrycks2019benchmarking} are adopted to reduce bias and improve fairness in the crime rate prediction problem. However, experimental results show that the mitigation methods are not always effective, which means the bias mitigation methods do not work in all real-world applications. By comparing the effects of these mitigation methods on different data splitting methods, different machine learning models, and different sensitive features, we find that these mitigation methods cannot reduce unfairness in some cases, and even aggravate it. The persistence of bias across different models and mitigation methods indicates that the problem is not algorithmic, but rooted in the data itself, and also highlights the risk of over-relying on technical solutions without understanding the context of the data. This suggests that bias mitigation approaches should be rigorously evaluated in each deployment context rather than assumed to be universally effective.

To understand the reasons for the mitigation failures in predictive models on government data, we first explore the potential sources of unfairness. There will be some policy changes in government operations, and the government data before and after the changes will shift in the feature space. Moreover, some emergencies, such as COVID-19 and lockdown, can also cause dramatic changes in data at the time, which can be reflected as data shift. Data shift makes bias mitigation difficult because mitigation methods usually require that the distribution of training data and test data is consistent. The accumulation of historical biases is another source of unfairness, which leads to implicit biases in the data. These historical biases are difficult to mitigate by adjusting a single model or a specific data set and require intervention from multiple aspects~\cite{davis2021algorithmic}. The lag in data release is also a challenge for bias mitigation.  For example, the census is conducted every ten years, and it is difficult to use the mitigation method to address the actual bias in the 2024 government data by using the 2021 census data. In addition, through a set of intersectional experiments by combinations of race and religion features, a problem attracts our attention. Using only a single sensitive feature for fairness analysis and mitigation may result in the potential systemic oppression of vulnerable groups.

Unlike prior studies that audit algorithmic decision-making in practice, our work focuses on understanding why mitigation failures occur, using a controlled setting grounded in real government data. This perspective complements existing literature on algorithmic policing by focusing not on whether models produce unfair outcomes, but on why widely used fairness interventions often fail when applied to real government data.
Therefore,  when designing AI for government services, we recommend that these factors should be considered in collaboration with social scientists who have a deep knowledge of the contexts in which these data are both drawn from and applied. The unfairness problem is not only a data and algorithm problem, but also a social, cultural and ethical issue. Solving the unfairness in AI for government service requires comprehensive consideration of historical background, social structure and group behaviour, and social scientists have deep expertise in these areas which suggests the need for an interdisciplinary approach to the consideration of unfairness in AI systems.

\section{Case Study: Crime Prediction in Bristol}
Crime prediction in Bristol is selected as the representative case study to examine the bias and its mitigation effectiveness in the public sector. This section presents an overview of the study, including  (1) the ML models implemented, and (2) model performance in predicting crime rates. The description and processing of the data used in this study are provided in Appendix C.  We provide code here~\footnote{https://github.com/HongboBo/Failing-on-Bias-Mitigation}.

\subsection{Regression Task for Crime Prediction}
\label{sec:regresstion}
In this study, crime prediction is defined as a regression task where the aim is to estimate the regression model \( f: \mathcal{X} \rightarrow \mathcal{Y} \), with \( \mathcal{X} \) representing the multidimensional feature data related to the Ward (a local area for voting purposes), and \( \mathcal{Y} \) representing the crime rate of a certain type of crime in a Ward at a specific year. To achieve the crime prediction model, we attempt the following regression models  \( f\): Multi-Layer Perceptron (MLP),  Decision Tree Regressor (DT), Random Forest Regressor (RF),  Gradient Boosting Regressor (GB), and a basic Linear Regression (LR) model. The implementation details are described in Appendix A.1.

While we do not use any deployed crime risk prediction systems, our modeling approach reflects the algorithms commonly used in real-world applications. For example, the Durham HART model is built on random forests~\cite{oswald2018algorithmic},  Chicago SSL uses boosted regression to estimate the violence propensity score~\cite{saunders2016predictions}, and COMPAS system has not been disclosed but simple algorithms have been shown to perform predictions approximately as well as the COMPAS~\cite{angelino2018learning}. These real-world systems tend to use relatively fundamental models, and we also adopt similar models to construct the experimental environment. By training them on publicly available government data,  our setup reflects the structure of predictive policing tasks, but avoids the complexity of real-world systems, which often rely on proprietary models and data which are difficult for us to obtain. Also, this study aims to analyze bias in the data rather than improve model accuracy, so simple and transparent traditional models are more suitable. Therefore, these models allow us to analyze bias mitigation in a controlled setting, where different models and mitigation methods can be systematically compared under consistent conditions.

For our experiments, we employed two different data splits on the pre-processed dataset:
 \begin{itemize}
     \item Temporal Split: The dataset is split based on the `year', with data from 2016 to 2021 used as the training set and data from 2022 as the test set, reflecting a real-world crime prediction scenario.
     \item Random Split: To facilitate the analysis of our findings during the bias mitigation process, the dataset is randomly split into 80/20\% for training and testing. This is a common data-splitting strategy for machine learning model training which can ensure the data distribution balance across both training and testing sets.
 \end{itemize}

The mean absolute error (MAE) and the coefficient of determination ($R^2$) are used as evaluation metrics to assess model performance. 
Let $D = \{ (x_i, y_i) \}_{i=1}^n$ denote the dataset to be evaluated, where $x_i$ represents the feature values and $y_i$ represents the actual crime rate for the $i$-th sample. Given the model's predicted crime rate for the $i$-th sample as $f(x_i)$, these metrics can be defined as follows, where all summations are over $i$ from $1$ to $n$:
 
\begin{equation}
\text{MAE}(D) = \frac{\sum |y_i - f(x_i)|}{n}
\end{equation}
\begin{equation}
    R^2 (D) = 1 - \frac{\sum(y_i - f(x_i))^2}{\sum (y_i - \frac{1}{n} \sum y_i)^2}
\end{equation}

\subsection{Experimental Results}

Table~\ref{tab:crime} presents the performance of different models on the processed testing sets.  Note that since the task is to predict the number of crimes per 1,000 people and the maximum value is 476.6, the $\mathcal{Y}$ ranges from 0 to 476.6. The values of MAE are mainly between 3 and 6 (except for linear regression), for example, the MAE of Random Forest (RF) is 3.49 (Temporal Split), accounting for 0.73\% of the target variable range; the MAE of MLP is 5.07, accounting for 1.06\%, and this gap may be negligible in practical applications, especially for low-noise tasks. The $R^2$ values of each model are around 0.9 (except LR), indicating that the models have little difference in their ability to explain the target variable. We can conclude that, except for LR, the other models perform well on data from two different split methods.
The LR model is excluded from the following fairness analysis due to its unsatisfactory performance on this task.

\begin{table}[htp]
\centering
\begin{tabular}{c|c|c|c|c}
\hline
\multirow{2}{*}{} & \multicolumn{2}{c|}{Temporal Split}      & \multicolumn{2}{c}{Random Split}      \\ \cline{2-5}
   & $MAE$ & $R^2$ & $MAE$ & $R^2$ \\ \hline
MLP  & 5.07$_{\pm0.22}$   & 0.91$_{\pm0.00}$  &  5.09$_{\pm0.13}$   & 0.90$_{\pm0.00}$   \\ 
DT   & 4.45$_{\pm0.08}$ & 0.93$_{\pm0.01}$ & 4.81$_{\pm0.10}$ & 0.89$_{\pm0.01}$ \\
RF   & 3.49$_{\pm0.03}$ & 0.94$_{\pm0.00}$ & 3.10$_{\pm0.03}$ & 0.91$_{\pm0.00}$ \\
GB   & 5.33$_{\pm0.00}$ & 0.92$_{\pm0.00}$ & 5.22$_{\pm0.01}$ & 0.90$_{\pm0.00}$ \\
LR   & $>$100 & $<$0 & 10.79$_{\pm0.69}$ & 0.66$_{\pm0.05}$ \\

\hline
\end{tabular}
\caption{Performance of Models on Temporal and Random Splits. Results are reported in mean $\pm$ standard deviation on 10 runs.}
\label{tab:crime}
\end{table}

\section{Fairness Analysis}
In this section, a series of experiments are conducted to examine the fairness of model prediction. Following these analyses, bias mitigation methods are adapted to reduce unfairness in model prediction performance. The data-level analysis, which provides insights into potential biases in sensitive features and crime rates, is presented in Appendix E.

\subsection{Model Fairness Analysis}
\label{sec:fairness}
We design a series of fairness analysis experiments to examine models' fairness performance under different sensitive features.

\textbf{Fairness Estimation Metrics.}
Most current model fairness studies focus on discrete sensitive features. For example, when the sensitive feature is gender, the data set can be divided into \textit{Male} and \textit{Female} groups, and the model's performance on each group can be analyzed to evaluate model fairness. The most common evaluation metric is to calculate the difference or ratio of the model performance in different subgroups~\cite{chouldechova2017fair, hu2024proximix, shah2022selective}.
Fairness issues with continuous sensitive features are an area that is underexplored. Here, we propose discretizing continuous sensitive features based on a certain threshold and convert the problem to the traditional fair regression discussion \cite{zhang2021assessing,agarwal2019fair}.
Let the sensitive feature class be \( C = \{C_1, C_2, C_3, \ldots, C_p\} \) representing categories like race and religion. For each class \( c \in C \), we define the feature set \( S_c = \{S_c^1, S_c^2, \ldots, S_c^k\} \), for example, \( S_{\text{race}} = \{\text{Black}, \text{White}, \text{Middle East}\} \). Each sensitive feature \( \phi \in S_c \) has a threshold \( T_{\phi} \), which is defined as the average of the maximum and minimum values of the given $\phi$. Discretization is applied by grouping the data to \( D_{\text{Low}[\phi]} \) if \( x_i(\phi) < T_{\phi} \), and to \( D_{\text{High}[\phi]} \) otherwise, where \(x_i(\phi)\) is the vaule of feature $\phi$ for the $i$-th sample. For instance, given a sensitive feature with continuous values (e.g., the proportion of \textit{White} between 0 and 1), we divide the samples with a proportion of \textit{White} below a threshold into the \( D_{\text{Low}[\textit{White}]} \) and other samples into the \( D_{\text{High}[\textit{White}]} \) . Then, a fair model should maintain consistent performance across different intervals of continuous feature values (such as high and low ranges of the proportion of \textit{White}).

To provide a robust and multidimensional evaluation of fairness, we assess model bias across \(S_{\text{race}}\) and \(S_{\text{religion}}\). We select six  features from \(S_{\text{race}}\) and six from \(S_{\text{religion}}\)  including major race and religion groups, key minorities, and also some very small groups, and the statistics are shown
in Figure~\ref{fig:race_stat} in Appendix E.  Additionally, our estimation involves both (1) absolute and (2) relative group fairness performance calculations:

(1) We calculate MAE for the model performance on each feature group (such as \( D_{\text{Low}[\phi]} \) or \( D_{\text{High}[\phi]} \)) respectively.

(2) We use the MAE disparity ($\Delta$MAE) as a measure of fairness to directly reflect the difference in prediction performance between different groups. Given a sensitive feature $\phi$, the $\Delta$MAE is defined as: 
\begin{equation}
\Delta \text{MAE} = \left| \text{MAE}(D_{\text{Low}[\phi]}) - \text{MAE}(D_{\text{High}[\phi]}) \right|
\end{equation}
$\Delta$MAE measures the difference in performance of the model on different groups of a sensitive feature. If $\Delta$MAE is close to 0, it means that the model performs consistently on both groups and the error distribution is fair.
If $\Delta$MAE is large, it means that the model has a significantly higher error on one group and there is performance imbalance or unfairness.

\begin{figure*}[t]
    \centering
    \includegraphics[width=\textwidth]{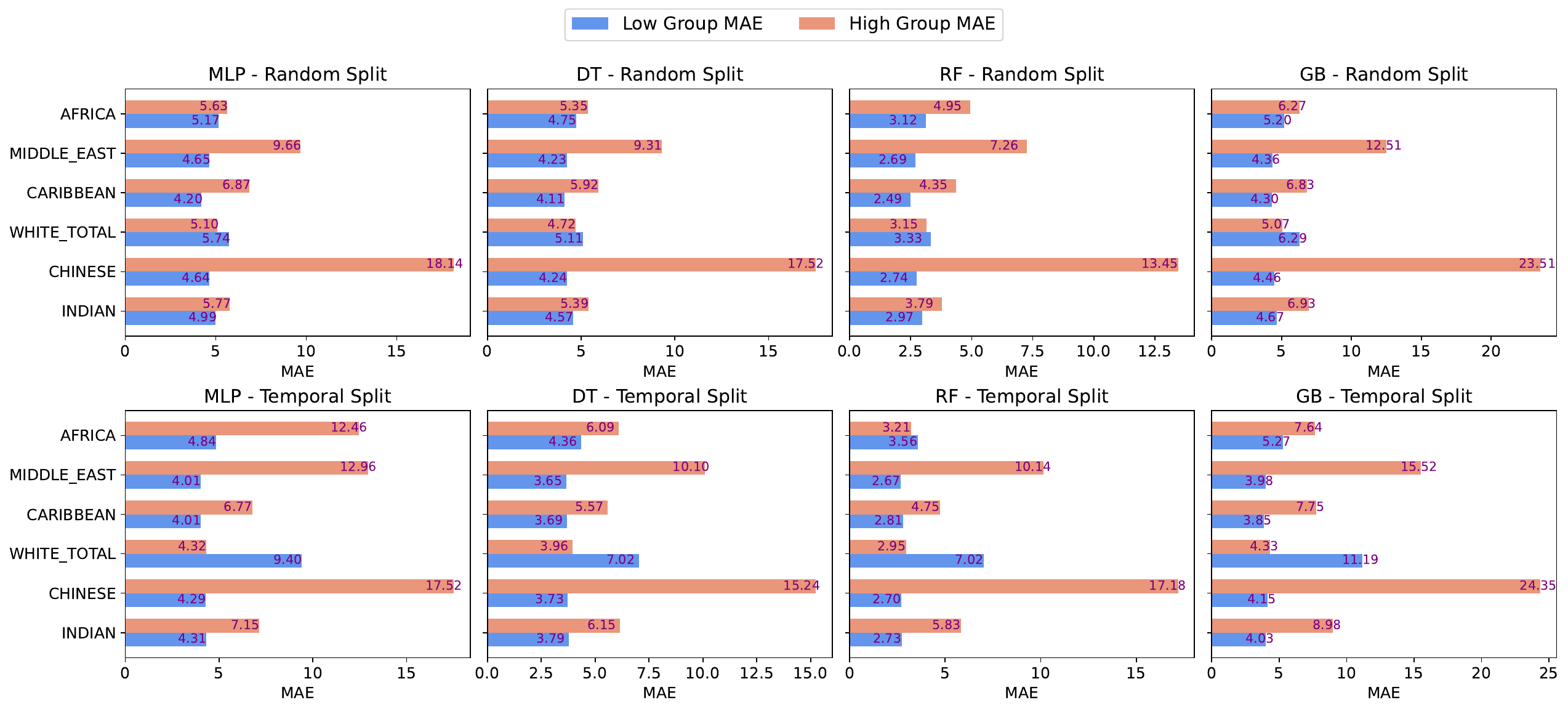}
    \caption{Single Feature Fairness Analysis. Each race feature is divided into low-high groups and tested on four different models. The difference between the High and Low groups of the same feature shows the unfairness when the model makes the prediction. The results in this figure are the average of 10 runs of experiments. The fairness analysis on religion features is shown in Figure~\ref{fig:single_rel} in Appendix B.1.}
    \label{fig:single}
\end{figure*}

\textbf{Results}
 The experimental results for race features fairness analysis are presented in Figure~\ref{fig:single}. For a model to be considered fair, the test errors between the Low group \(D_{\text{Low}[\phi]}\) and High group \(D_{\text{High}[\phi]}\)  of a single feature $\phi$ should be comparable. However, we can see that the four models are unfair in both data splits. For race, the models generally perform poorly on the Wards with a low proportion of \textit{white} people and Wards with a high proportion of minority ethnic, and the unfairness is most obvious on the Wards with a high proportion of \textit{Chinese} people. 
The corresponding analysis for the religion features is provided in Figure~\ref{fig:single_rel} in Appendix B.1, where unfairness issues are also observed. To further analyse this unfairness, we refer to the slope lines illustrated in Figure~\ref{fig:race} (presented in Appendix E). A steeper slope of the red line usually shows greater unfairness in Figure~\ref{fig:single}. Moreover, if the red line shows an upward trend, then the unfairness is concentrated in the High group, whereas a downward trend indicates unfairness in the Low group.

\subsection{Bias Mitigation Strategies}

We adapt and refine four classical bias mitigation methods that cover both data- and model-level strategies, commonly applied in fairness research for tabular data.  These methods are applied to address issues of unfairness associated with the sensitive feature $\phi$ in population data, and the implementation details are described in Appendix A.2.

\begin{itemize}
    \item OverSampling (OS)~\cite{he2009learning}: Resample the sample to ensure that the number of samples between \( D_{\text{Low}[\phi]} \) and \( D_{\text{High}[\phi]} \)) is balanced. In this study, random oversampling is used to obtain the duplicate data within the smaller sample sized group.
    \item MixUp (MU)~\cite{zhang2017mixup}: By generating new samples, unsampled areas in the feature space are simulated while preserving the statistical properties of the original data. In this study, a group-wise MixUp strategy is used to generate the sample. 
    \item Perturbation (Pert.)~\cite{hendrycks2019benchmarking}: During the resampling process, Gaussian noise is added to the feature values of the target samples. Adding noise can enhance data diversity and reduce bias in the model by avoiding the repeatedly generated samples.
    \item ReWeight (RW)~\cite{zafar2017fairness}: Reweighting is a commonly used bias mitigation technique that aims to reduce the bias of the model towards a specific group by adjusting the weights of samples in the training data.

\end{itemize}

\begin{table*}[t]
\centering
\setlength{\tabcolsep}{1mm}

{\fontsize{9pt}{11pt}\selectfont
\begin{tabular}{c|c|c|c|c|c|c|c|c|c|c|c}
\hline
\multirow{2}{*}{} & \multirow{2}{*}{Model} & \multicolumn{5}{c|}{Temporal Split}      & \multicolumn{5}{c}{Random Split}  \\ \cline{3-12}
  &  & \rule{0.5cm}{0.2mm} & OS & MU & Pert. & RW & 
    \rule{0.5cm}{0.2mm} & OS & MU & Pert. & RW  \\ \hline
\multirow{4}{*}{Chinese}
& MLP & 13.23$_{\pm1.30}$ & 13.78$_{\pm2.38}$ & 15.00$_{\pm3.66}$ & 24.39$_{\pm7.74}$ & 14.55$_{\pm2.67}$
      & 13.50$_{\pm0.59}$ & \underline{9.35}$_{\pm3.10}$ & \underline{9.27}$_{\pm3.96}$ &10.91$_{\pm3.22}$ & 10.78$_{\pm3.51}$\\
& DT  & 11.51$_{\pm1.28}$ & 12.53$_{\pm1.07}$ & 15.12$_{\pm2.10}$ & 13.04$_{\pm1.33}$ & 45.41$_{\pm7.37}$
      & 13.28$_{\pm2.63}$ & 13.72$_{\pm4.55}$ & 15.88$_{\pm6.53}$ & 18.19$_{\pm5.31}$ & 14.33$_{\pm4.25}$\\
& RF  & 14.48$_{\pm1.16}$ & 15.13$_{\pm1.63}$ & \underline{9.77}$_{\pm0.67}$ & 14.25$_{\pm1.65}$ & 20.34$_{\pm1.08}$ 
      & 10.70$_{\pm0.48}$ & 11.85$_{\pm4.06}$ & 12.02$_{\pm4.44}$ & 11.59$_{\pm3.91}$ & 13.04$_{\pm3.67}$\\
& GB  & 20.20$_{\pm0.16}$ & 19.15$_{\pm0.20}$ & 18.18$_{\pm0.82}$ & 17.67$_{\pm2.25}$ & 18.55$_{\pm0.25}$ 
      & 19.04$_{\pm0.27}$ & \underline{13.77}$_{\pm3.61}$ & 16.25$_{\pm4.25}$ & 14.87$_{\pm3.39}$ & \underline{14.20}$_{\pm3.66}$\\ \hline

\multirow{4}{*}{\shortstack{Middle\\East}}
& MLP & 8.95$_{\pm0.95}$ & 7.07$_{\pm0.72}$ & 9.10$_{\pm2.30}$ & 14.97$_{\pm7.94}$ & \underline{6.59}$_{\pm0.60}$ 
      & 5.01$_{\pm0.25}$ & 5.27$_{\pm1.3}$ & 5.81$_{\pm2.00}$ & 5.81$_{\pm2.02}$ & 5.09$_{\pm1.37}$\\
& DT  & 6.45$_{\pm0.84}$ & 5.26$_{\pm0.86}$ & 7.74$_{\pm0.93}$ & 5.35$_{\pm1.10}$ & 4.98$_{\pm0.95}$
      & 5.08$_{\pm1.05}$ & 7.84$_{\pm2.57}$ & 7.95$_{\pm2.35}$ & 8.24$_{\pm3.11}$ & 7.14$_{\pm1.80}$ \\
& RF  & 7.47$_{\pm0.57}$ & 9.91$_{\pm1.27}$ & \underline{5.54}$_{\pm0.40}$ & 8.02$_{\pm0.81}$ & 9.85$_{\pm0.98}$ 
      & 4.57$_{\pm0.14}$ & 6.15$_{\pm2.2}$ & 6.32$_{\pm1.94}$ & 6.27$_{\pm2.23}$ & 6.71$_{\pm2.37}$ \\
& GB  & 11.53$_{\pm0.10}$ & 11.97$_{\pm0.13}$ & 11.88$_{\pm0.63}$ & 9.55$_{\pm2.24}$ & 11.58$_{\pm0.13}$ 
      & 8.14$_{\pm0.10}$ & 7.34$_{\pm2.02}$ & 8.64$_{\pm2.37}$ & 8.24$_{\pm2.08}$ & 7.35$_{\pm2.07}$\\ \hline

\multirow{4}{*}{Christian}
& MLP & 2.13$_{\pm0.54}$ & 2.09$_{\pm0.35}$ & 2.43$_{\pm0.39}$ & 3.17$_{\pm1.23}$ & 2.15$_{\pm0.76}$ 
      & 2.25$_{\pm0.66}$ & 2.29$_{\pm1.38}$ & 2.05$_{\pm0.67}$ & 2.20$_{\pm1.26}$ & 2.39$_{\pm0.74}$\\
& DT  & 1.93$_{\pm0.29}$ & 1.92$_{\pm0.19}$ & 1.75$_{\pm0.15}$ & 2.21$_{\pm0.39}$ & 1.82$_{\pm0.19}$
      & 2.66$_{\pm1.12}$ & 2.49$_{\pm0.97}$ & 2.45$_{\pm1.17}$ & \underline{1.75}$_{\pm1.05}$ & 2.55$_{\pm1.00}$ \\
& RF  & 1.86$_{\pm0.16}$ & 1.73$_{\pm0.11}$ & 1.77$_{\pm0.09}$ & 1.45$_{\pm0.24}$ & 1.85$_{\pm0.12}$ 
      & 1.87$_{\pm0.96}$ & 1.83$_{\pm1.32}$ & 1.92$_{\pm0.96}$ & 1.83$_{\pm1.56}$ & 1.89$_{\pm0.95}$\\
& GB  & 3.44$_{\pm0.03}$ & 3.88$_{\pm0.04}$ & 3.41$_{\pm0.13}$ & 3.30$_{\pm0.13}$ & 3.44$_{\pm0.02}$ 
      & 2.74$_{\pm0.99}$ & 2.86$_{\pm1.21}$ & 2.75$_{\pm1.08}$ & 2.80$_{\pm1.31}$ & 2.73$_{\pm1.02}$\\ \hline

\multirow{4}{*}{Buddhist}
& MLP & 3.59$_{\pm0.69}$ & 5.14$_{\pm0.55}$ & 4.91$_{\pm1.03}$ & 10.45$_{\pm2.09}$ & 3.99$_{\pm0.54}$ 
      & 4.75$_{\pm0.98}$ & 4.92$_{\pm1.66}$ & \underline{3.56}$_{\pm1.24}$ & 4.29$_{\pm1.06}$ & 4.51$_{\pm1.01}$\\
& DT  & 3.65$_{\pm0.54}$ & 4.89$_{\pm0.52}$ & 4.40$_{\pm0.90}$ & 5.39$_{\pm0.87}$ & 4.21$_{\pm0.59}$ 
      & 4.76$_{\pm1.31}$ & 4.82$_{\pm1.36}$ & 4.42$_{\pm1.62}$ & 4.15$_{\pm1.32}$ & 4.85$_{\pm1.53}$\\
& RF  & 4.25$_{\pm0.30}$ & 3.54$_{\pm0.15}$ & 3.61$_{\pm0.41}$ & 4.94$_{\pm0.44}$ & 5.02$_{\pm0.51}$ 
      & 3.52$_{\pm1.48}$ & 3.52$_{\pm1.25}$ & 2.90$_{\pm1.08}$ & 3.76$_{\pm1.05}$ & 3.53$_{\pm1.51}$\\
& GB  & 6.68$_{\pm0.05}$ & 5.61$_{\pm0.04}$ & 6.77$_{\pm0.67}$ & 6.52$_{\pm0.33}$ & 6.79$_{\pm0.05}$ 
      & 5.24$_{\pm1.26}$ & 5.51$_{\pm1.25}$ & 5.23$_{\pm1.34}$ & 6.25$_{\pm1.49}$ & 5.08$_{\pm1.16}$\\

\hline
\end{tabular}
}
\caption{Average Bias Mitigation Performance (Evaluated by $\Delta$MAE). `--' means no mitigation method is used which is the baseline. We underline the \underline{values} of methods that achieve a mitigation performance improvement exceeding 25\% over the baseline. Each result is reported in mean on 10 runs and the lower the value, the better the effect.}
\label{tab:mitigation}
\end{table*}

The experimental results are shown in Table~\ref{tab:mitigation} and the results of mitigation methods with the improvement exceeding 25\% over the baseline are marked as `effective'. The results indicate that not all mitigation methods are effective, and not all sensitive features in the dataset can be successfully mitigated. Specifically, under different data splitting methods, significant improvements are observed in only three results using the Temporal Split data, compared to six results with the Random Split data, which shows it is easier to perform bias mitigation on Random Split data. Regarding mitigation methods, OverSampling produced two effective results, MixUp achieved four, Perturbation resulted in one, and ReWeight produced two, which shows different mitigation methods could be effective in specific data splits or models; For different machine learning models, four effective results are observed for MLP, while Decision Tree, Random Forest, and Gradient Boost each achieved two effective results, which shows the models that are less dependent on data distributions can be more effective to be mitigated.

\section{Discussion}

In this section, we start with a discussion about the potential sources of unfairness caused by the government data used in this case study. Next, we explore the blind spots in fairness analysis through a set of interactional fairness experiments. Finally, we discuss the implications of these findings and provide recommendations for addressing unfairness in AI models designed to support government services.

\subsection{Potential Sources of Unfairness}

Given the challenges in mitigating the unfairness of the models trained on government data, we aim to explore the potential sources of unfairness. After testing of multiple models and mitigation methods, bias persisted, indicating that the root cause should lie in the data itself -- models learn from data and the underlying bias in the data will be learned and reproduced by the models.  We observe three features in government data: they often evolve with policy changes or emergencies, reflect accumulated historical biases, and may suffer from delays in information release. These factors respectively lead to data distribution shifts, systemic unfairness, and a failure to reflect current realities, making it especially challenging to mitigate unfairness effectively.

\subsubsection{Data distribution shift}
\begin{figure}[t]
    \centering
    \includegraphics[width=\linewidth]{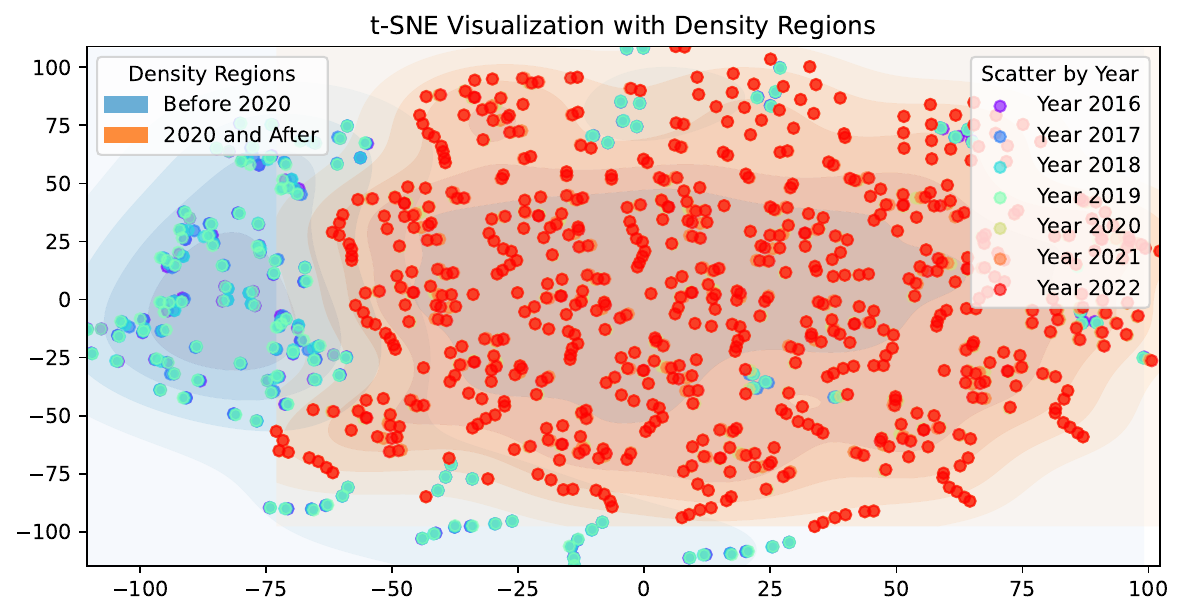}
    \caption{Scatter Plot of Data in Each Year. The data samples are visualized by t-SNE~\cite{van2008visualizing}. The feature distribution before (area in blue) and after 2020 (area in red) has significantly shifted in the feature space. }
    \label{fig:distribution}
\end{figure}

We first demonstrate the phenomenon of shift in data, and then discuss the policy changes and emergencies that cause data shift.
Government data is usually affected by data changes over time, and data distribution is not static and uniform. As shown in Figure~\ref{fig:distribution}, we can see that in different years, the distribution of data in latent feature space has shifted. We quantify the shift between 2016 and 2022 by the Maximum Mean Discrepancy~\cite{gretton2012kernel}, which yielded a value of 0.106 in the feature space, and 57.63\% of features having a significant shift (p $<$0.05). This distribution shift makes bias mitigation unstable in the Temporal Split scenario, study~\cite{shao2024supervised} has also focused on this challenge of bias mitigation on data distribution shift. However, the same bias mitigation methods have more effects on models trained under Random Split data. That is because the distribution of Random Split data is more even and the mitigation methods, like OverSampling and MixUp, can effectively expand the number of training sample spaces of a few groups when the data distribution is the same, and then address the unfairness issue.

\begin{figure}[t]
    \centering
    \begin{subfigure}[b]{0.45\textwidth}
    \centering
    \includegraphics[width=\textwidth]{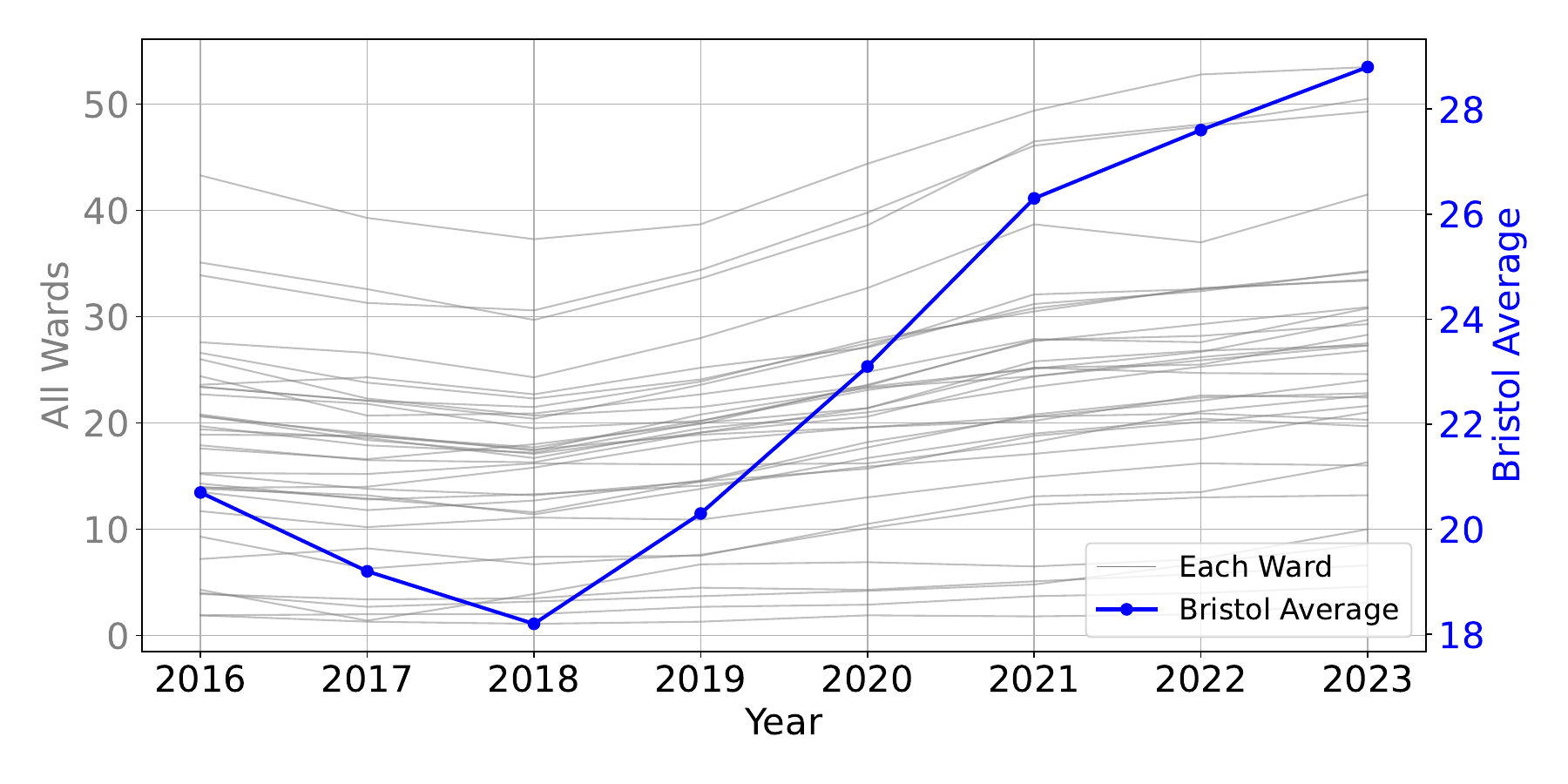}
    \caption{Free School Meals Statistics. The values on the y-axis are the rate (\%) of the pupils who are eligible for free school meals. }
    \label{fig:meals}
    \end{subfigure}
    \begin{subfigure}[b]{0.45\textwidth}
    \centering
    \includegraphics[width=\textwidth]{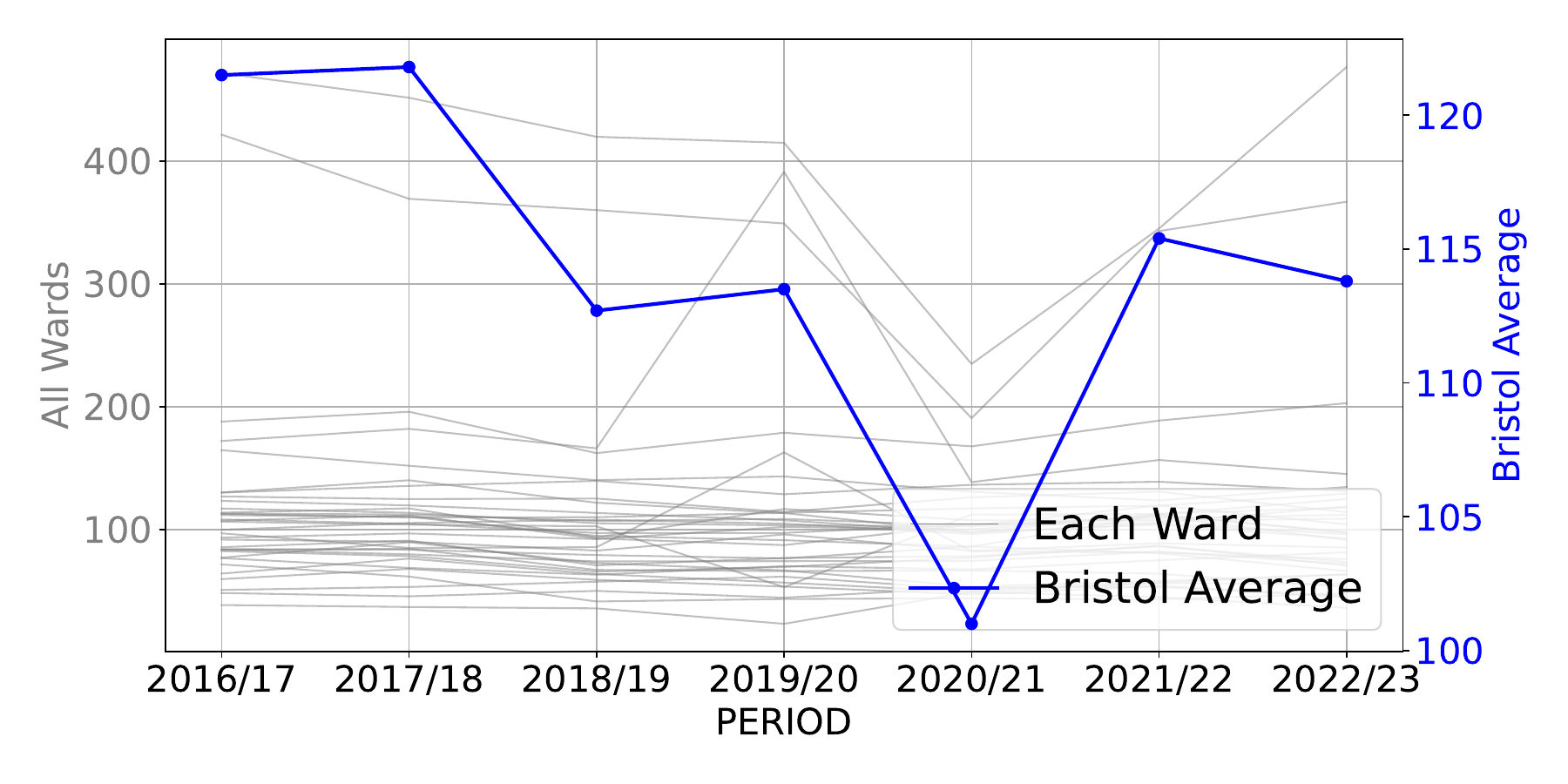}
    \caption{Crimes Statistics. The values on the y-axis are the crime rate (number per 1000 population).}
    \label{fig:crime}
    \end{subfigure}
    \caption{The grey lines (left y-axis) represent individual Wards, while the blue line (right y-axis) highlights the Bristol Average. The purpose of using a dual-axis plot is to highlight the average trend in which (a) the requirement for free school meals suddenly increased after 2018 and (b) the crime rate showed a significant decrease in the COVID lockdown (2020/21) period.}
    
\end{figure}

\textbf{Unpredictability of policy changes} Changes in government policies are often intended to address specific social issues. However, bias mitigation techniques are mostly based on historical trends of data features and cannot predict the change brought about by a new policy. We observe that the trends of certain features in government data can suddenly change, as illustrated in Figure~\ref{fig:meals}, which shows a trend change in free school meals used in this study. Such shifts increase the unpredictability of feature variations. We find that in April 2018, the UK Department for Education updated the eligibility criteria~\cite{DfE2024FSMGuidance} for free school meals to ensure more pupils in need can benefit from free school meals. We assume that this data change may be related to the increase in the amount of free meal since 2018. In addition to policy changes, changes in the social or economic environment may also cause unpredictable data changes. For example, economic inactivity in Bristol would affect employment rates~\cite{ONS2024BristolLabourMarket}, but we do not use this data in this study so we will not discuss this. 
This makes it challenging for some mitigation methods, such as MixUp, to generate features that can cover the distribution of the test set by mixing the training set, because these unpredictable changes will lead to unpredictable data distribution changes.

\textbf{Impact of emergencies.} The AI models for supporting government decisions are usually trained on historical data, but when emergencies change the citizen's behaviour, these historical data lose their value for referencing. For example, the government data used in this study was affected by changes in COVID-19. During the COVID pandemic, the crime rate shows a significant decrease due to the lockdown policy~\cite{ukcrimecovid2021}, as shown in Figure~\ref{fig:crime}. Unlike data trend changes driven by policy changes, such as the downward trend before 2018 and the upward trend after 2018 in free school meal data, changes triggered by emergent events typically manifest as outliers during the event period. Once the event ends, the data generally returns to its normal level. This creates anomaly data, further exacerbating the challenges of crime prediction and bias mitigation. Many bias mitigation methods, especially reweighting or resampling techniques, assume that the data comes from the same distribution. However, the changes caused by emergencies are often drastic and nonlinear, which will undermine the assumption of the same distribution, resulting in the ineffectiveness of mitigation methods. Therefore, the core challenge brought by emergencies is still data distribution shift.

\subsubsection{The accumulation and reproduction of historical and structural biases}

We remove all the race and religion features from the training set and re-perform the model fairness analysis experiments. We find that even if there were no sensitive features in the training set, the trained models are still unfair on the sensitive race and religion features, as shown in Table~\ref{tab:no_sensitive}. This occurs because, although sensitive features are excluded as inputs, other variables, such as information related to pupils and jobs, indirectly reflect the distribution of race and religion. Since part of the information in government data is closely related to structural social inequality, what the model `learns' is actually the accumulation of bias from historical evolution, which will continue to be reproduced over time~\cite{davis2021algorithmic}. 
Even when models appear `fair' by excluding sensitive features, they can still contain implicit biases. Therefore, simply omitting sensitive information, like race and religion, doesn't guarantee fairer models.

\begin{table*}[t]

\setlength{\tabcolsep}{1mm}
\begin{tabular}{c|cccccc|cccccc}
\hline
       & \multicolumn{6}{c|}{Race-related Sensitive Features}                 & \multicolumn{6}{c}{Religion-related Sensitive Features}             \\
 & Indian & Chinese & White & Middle East & Caribbean & Africa & Muslim & Jewish & Hindu & Buddhist & Christian & No Religion \\
\hline
With     & 2    & 13.43 & 2.52 & 2.72 & 7.11 & 0.04 & 0.91 & 1.46 & 13.86 & 4.66 & 2.52 & 1.72 \\
\hline
Without & 0.78 & 13.5  & 0.64 & 2.67 & 5.01 & 0.46 & 0.42 & 1.71 & 12.62 & 4.76 & 2.26 & 1.26 \\
\hline
$|\text{Diff}|$     & 1.22 & 0.07  & 1.88 & 0.05 & 2.1  & 0.42 & 0.49 & 0.25 & 1.24  & 0.1  & 0.26 & 0.46 \\
\hline
\end{tabular}
\caption{Models' Fairness Performance Comparison Between Training With vs Without Sensitive Features (Bias evaluated by \(\Delta\)MAE, under MLP with Random Split; the full results are in Figure \ref{fig:no_sensitive} in Appendix B.2). The differences $|\text{Diff}|$ are slight, where unfairness persists in the sensitive feature groups even after the sensitive features have been removed.}
\label{tab:no_sensitive}
\end{table*}

\subsubsection{Data release lags}
Government data, especially identity-related features (e.g., race, religion), often comes from large-scale population censuses, which are generally conducted once a decade. These less frequently updated features are far more lagging compared to annually refreshed data such as regional crime rates. These lags prevent data from reflecting how policies or emergencies alter population characteristics during the decade. It's hard for the model to find and mitigate potential unfairness effectively when the population structure has already shifted but models still rely on obsolete statistical data. Besides advocating for governments to release data in a more transparent and timely manner, one technical solution would be to develop additional models to simulate the dynamic changes of the populations. Mitigating the delays in statistics also aligns with the vision of achieving long-term fairness \cite{liu2018delayed}.

\subsection{The Blind Spots in Fairness
Analysis: Intersectional Fairness Perspective}

Besides the characteristics of government data, the current mainstream single-feature-based fairness evaluation in the ML community also deserves to be re-examined. The single-feature fairness analysis requires examining bias performance on different sensitive features (e.g., race, and religion) separately, which ignores the complex reality where these features intersect and interact. For instance, \textit{Muslims} are widely present in the populations of the \textit{Middle East} and \textit{Africa}. If a single feature fairness analysis is conducted on the \textit{Muslim} population without considering race, it would be difficult to determine whether \textit{Middle Eastern Muslims} would present the same results as \textit{African Muslims}. To explore further how the complex relationships between sensitive features can affect unfairness variations, we perform intersectional fairness analysis. Detailed experimental setup, formulae, and results are presented in Appendix D.  The result indicates that the prevalent practice of evaluating protected features in isolation creates dangerous blind spots: models that appear fair when evaluated along a single dimension may contain significant biases. By relying solely on traditional single-feature fairness metrics, we risk implementing bias mitigation strategies that leave vulnerable intersectional populations behind.  We are potentially reinforcing discriminatory practices by declaring systems `fair' based on incomplete evaluation criteria.

\subsection{Implications and Recommendations}
We chose Bristol data because it covers multiple dimensions related to public governance, such as crime, population, and religion, and has the typical characteristics of real-world government data.
Although our analysis is based on a single dataset, the challenges we identify are not unique to our case, also been observed in other data-driven decision contexts~\cite{singh2021fairness,gupta2024trial,zuo2024understanding}.
 While we do not claim universal applicability of our findings, the consistency of mitigation failures across different models and fairness methods in our setting points to mechanisms that are likely relevant in similar contexts. And these findings contribute to an early warning that biases in government data may persist even with standard mitigation methods.

For researchers in data science and AI, it is difficult to use existing mitigation methods to address unfairness in AI models designed for supporting government decision-making. This is because real society is complex and constantly evolving,  it is difficult to know the shift in future data distribution only using the information provided by the current data. Therefore, the models for government services should be constantly re-evaluated to ensure they are fair to use. Blindly applying fairness solutions that worked in one case to all AI models is risky, as the effective bias mitigation methods in some scenarios can fail in others.It is important to consider the impact of external factors on government data, such as how new policies or emergencies will affect society and how this will affect changes in government data. 

Understanding the impact of such factors and the ways in which unfairness operates in government data systems is best done in collaboration with social scientists who have a deep knowledge of the contexts in which these data are both drawn from and applied. Cross-disciplinary collaboration could support nuanced understandings of societal inequalities and the intersectional ways that inequalities operate to create structural disadvantages amongst the population which can be exacerbated in algorithmic systems that are trained on historical data which reflects these inequalities. While the field of AI fairness and the related subfield of fair machine learning are premised on resolving this problem of inequality reproduction. However, such efforts fall short because they are rooted in a false assumption of baseline meritocracy, achievable by neutralizing social differences within datasets and through algorithmic models~\cite{davis2021algorithmic}.  For example, social scientists can help AI researchers avoid missing the broader context with the information necessary to create more equitable outcomes by co-designing fairness metrics~\cite{selbst2019fairness} or developing proactive reparative mitigation strategies~\cite{davis2021algorithmic} based on interpreting how historical and structural biases accumulate.

\section{Conclusion}

AI models can help in government decision-making, but the unfairness that AI bring also needs attention. By using a crime prediction case, we find that some unfairness in models which are trained with government data is difficult to mitigate with technical means. Through further analysis, we find that the presence of data distribution shifts in government data, the accumulation of deep historical and structural biases, and the data failure to reflect current reality make it difficult for the unfairness of models to be mitigated. 
We also point out the limitations of the current mainstream fairness methods that use the single sensitive feature. That is, data considered fair with a single sensitive feature may not be fair when multiple sensitive features intersect.
The findings of this study provide a new insight on the application of AI in public service using government data, while also highlighting the challenges of fairness issues in the specific context of government data. Our work not only reveals the limitations of current mitigation methods but also highlights the importance of incorporating real-world society factors into the development of novel mitigation approaches specifically for government data in future research and practice.

\section{Acknowledgments}
The support of the Economic and Social Research Council (ESRC) is gratefully acknowledged. Grant Ref ES/W002639/1. Hongbo is funded by ESRC Centre for Sociodigital Futures (ES/W002639/1), Jingyu is funded by EPSRC-DTP ( EP/W524414/1/2894964). Debbie and Weiru are both partially funded by ESRC Centre for Sociodigital Futures (ES/W002639/1).

\bibliography{bibfile}

\clearpage

\appendix

\section{Appendix A: Implementation Details}
\subsection{A.1: Regression Models}

We provide implementation details for the regression models used in the section Regression Task for Crime Prediction.

\begin{itemize}
    \item A 2-layer Multi-Layer Perceptron (MLP): a learning rate of 0.01, trained with early stopping, MSE loss, and the Adam~\cite{kingma2014adam} optimizer;
    \item A Decision Tree Regressor (DT): maximum depth 10;
    \item A Random Forest Regressor (RF): 100 estimators;
    \item A Gradient Boosting Regressor (GB): 100 estimators, learning rate 0.1, and max depth 3; 
    \item A basic Linear Regression (LR) model;
\end{itemize}
DT, RF, GB, and LR are implemented using scikit-learn (v1.5.2), with unspecified parameters set to their default values. We also used several Python libraries for data processing and modeling, including NumPy (2.0.0), Pandas (2.2.3), and PyTorch (2.3.1), and conducted our experiments on a machine equipped with an Apple M3 Pro chip, running macOS. .

\subsection{A.2: Bias Mitigation Methods}
\label{apx:details}
To ensure transparency and reproducibility, we provide implementation details for the bias mitigation techniques used in our experiments. All methods were developed using \textit{Python} and implemented with standard data processing techniques like \textit{pandas} and \textit{numpy}, using the same experimental infrastructure described in Appendix A.1. A sensitive feature $\phi$ from  \(S_{\text{race}}\) or \(S_{\text{religion}}\) is consistently binned into low or high group (\( D_{\text{Low}[\phi]} \) or \( D_{\text{High}[\phi]} \)) based on the threshold \( T_{\phi} \). Below, we describe the procedures used for each method:

\begin{itemize}
    \item OverSampling (OS): We perform random oversampling by replicating samples from the underrepresented group within each label class until numerical parity is achieved. Sampling is done independently within each class to preserve class distribution and avoid label leakage.
    \item MixUp (MU): For each pair of randomly selected samples within the same sensitive group $(x_1, y_1)$ and $(x_2, y_2)$, we generate synthetic samples using the MixUp technique:\[
\tilde{\mathbf{x}} = \lambda \mathbf{x}_1 + (1 - \lambda) \mathbf{x}_2, \quad
\tilde{y} = \lambda y_1 + (1 - \lambda) y_2\] where $\quad \lambda \sim \mathrm{Beta}(\alpha, \alpha)$.

    \item Perturbation (Pert.): Gaussian noise with standard deviation $\sigma = 0.01$ is added to numerical feature columns of oversampled instances to prevent overfitting from duplicate data. This method enhances sample diversity while maintaining statistical coherence.
    \item ReWeighting (RW): We calculate adaptive sample weights inversely proportional to group frequencies after binning. Let $n$ denote the total number of samples and $n_g$ the number of samples in group $g$. The group weight is defined as: \[w_g=n/2\cdot n_g\] Each individual sample is assigned the weight $w_i = w_{g(i)}$, where $g(i)$ is the group to which sample $i$ belongs. We normalize the weights such that the maximum value is 1: \[w_i=w_{g(i)}/max(w_g)\]
    
    These weights are normalized and applied to each sample during loss computation, ensuring that underrepresented groups have greater influence during trainings.
\end{itemize}

\section{Appendix B: Additional Experiment Results}

\subsection{B.1: Fairness Analysis Results on Religion Features}
\begin{figure*}[t]
    \centering
    \includegraphics[width=\textwidth]{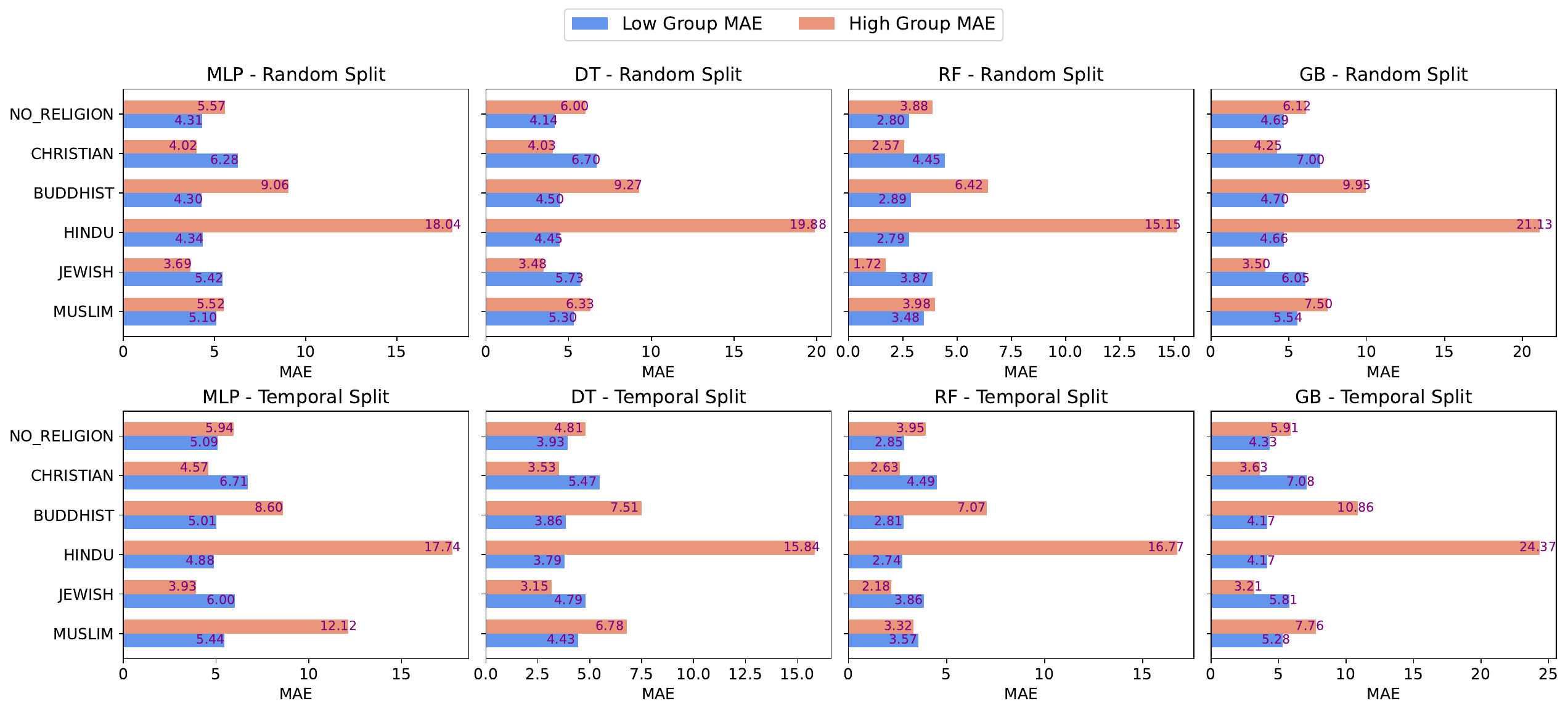}
    \caption{Single Feature Fairness Analysis on Religion Features. }
    \label{fig:single_rel}
\end{figure*}

Figure~\ref{fig:single_rel} presents the results of the single feature fairness analysis on religion features. Religion features are divided into low-high groups and tested on four different models. The difference between the high and low groups of the same feature shows the unfairness when the model makes the prediction.
The results in this figure are the average of 10 runs of experiments. 

\subsection{B.2: Fairness Performance Without Sensitive Features}
\label{apx:res}

\begin{figure*}[t]
    \centering
    \includegraphics[width=\linewidth]{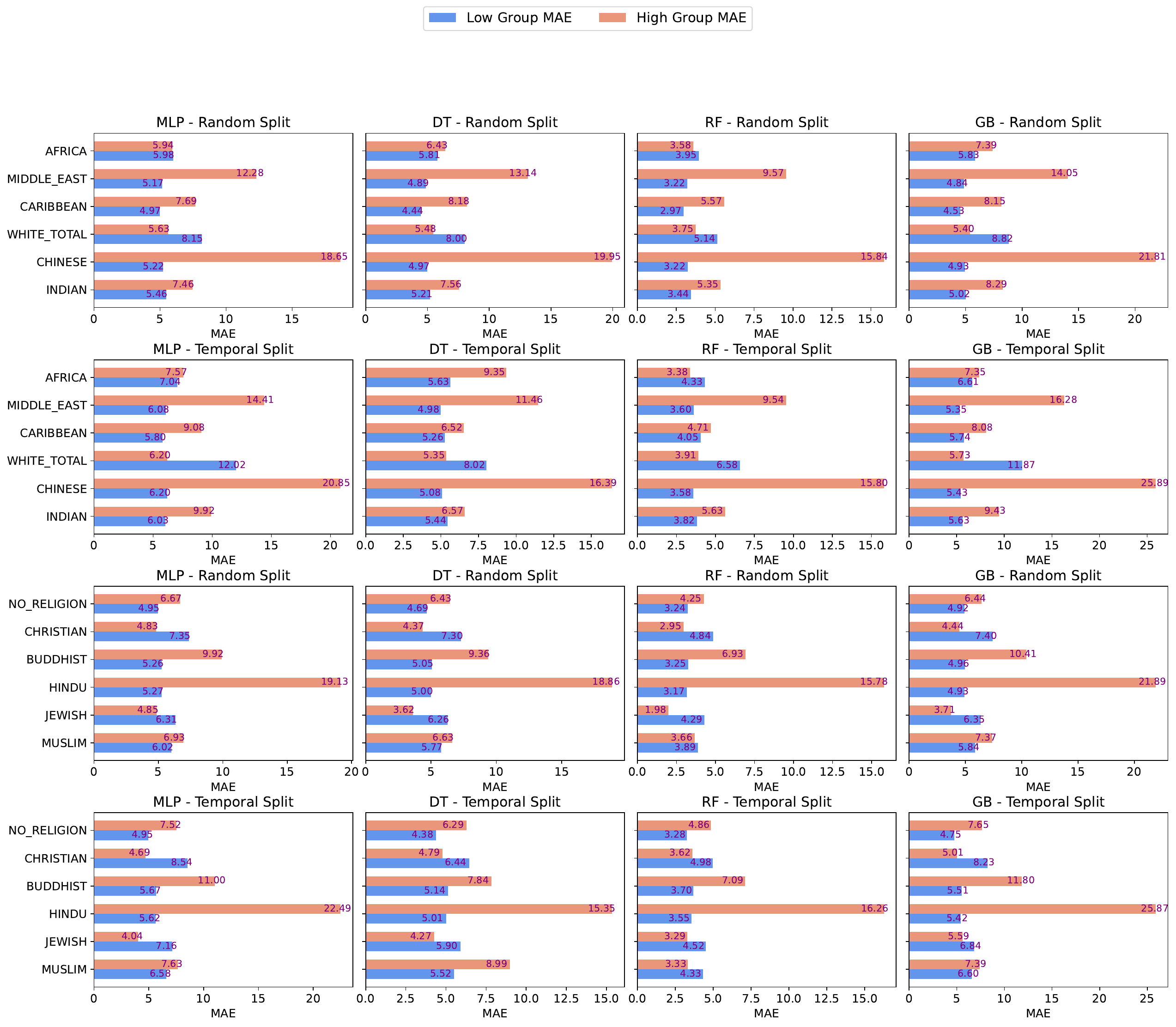}
    \caption{Single Feature Analysis without Sensitive Features Inputted during Training. There was still an unfairness in the sensitive feature groups in different machine learning models, even if the sensitive features were removed during training.}
    \label{fig:no_sensitive}
\end{figure*}

Figure~\ref{fig:no_sensitive} presents the complete results of the single feature fairness analysis without sensitive features inputted during training, serving as a supplement to the corresponding Table~\ref{tab:no_sensitive}.

\section{Appendix C: Data}
In this subsection, we describe the data from the \textit{Open Data Bristol} and how we process it for the following criminal statistics prediction task.

The data used in this study are from public datasets provided by the \textit{Open Data Bristol} platform~\footnote{\url{opendata.bristol.gov.uk}}, including the following topics: Identity, Jobs, Crime and Pupils. These data are locally held and administered by the city council. These data cover key indicators in different fields and are divided according to Ward~\footnote{A Ward is the local area for voting purposes.} level, reflecting the differences and distribution characteristics of the people living in the 34 different Wards in Bristol in various fields. 
\begin{itemize}
    \item Identity~\footnote{\url{opendata.bristol.gov.uk/datasets/identity-in-bristol-census-2021-by-ward/explore}}: The identity data is derived from the 2021 census and divided into Wards. It records the social and cultural characteristics of the Bristol population, including information such as race, religion, and language use. \footnote{In this study, we only focus on a selected subset of sensitive features.  We clarify that the selection does not reflect any assumption, it is primarily due to space limitations. We fully respect and encourage the rich diversity of races and religions beyond those included.}
    \item Crime~\footnote{\url{opendata.bristol.gov.uk/datasets/42eed1ee664c47309396756d05ac19ee_6/explore}}: The crime data provides crime statistics by Ward in Bristol from 2016/17 to 2022/23, covering the crime rate (numbers per 1000 population) and distribution of different types of crime.
    \item Jobs~\footnote{\url{opendata.bristol.gov.uk/datasets/bcc::jobs-by-industry-in-bristol-by-ward/explore}}: This data is divided into Wards and records the number of employees employed by the main industry plus working owners in Bristol from 2015 to 2022.
    \item Pupils Classed as Disadvantaged~\footnote{\url{opendata.bristol.gov.uk/datasets/bcc::pupils-classed-as-disadvantaged-in-bristol-by-ward-1/explore}; Disadvantaged pupils are defined by the Department for Education through eligibility for the Pupil Premium.}: This data details the number and rate(\%) of pupils in Bristol who are classed as disadvantaged, broken down by Ward from 2016 to 2023.
    \item Pupils Receiving Free School Meals~\footnote{\url{opendata.bristol.gov.uk/datasets/bcc::pupils-receiving-free-school-meals-in-bristol-by-ward-1/explore}}: This data is divided into Wards and its records in detail the number and rate(\%) of students who are eligible for receiving free school meals in each Ward in Bristol from 2016 to 2023.
    \item Pupils Absence~\footnote{\url{opendata.bristol.gov.uk/datasets/17a4eee3fbc44675ba84adfd3792c41d_24/explore}}: This data is divided by Ward and its records the absence rate of pupils in each Ward in the city of Bristol from 2016/17 to 2021/22. We have supplemented the 2022/23 data using the Open Data Bristol dashboard API~\footnote{\url{opendata.bristol.gov.uk/pages/dashboards}}.
    \item Youth Offenders~\footnote{\url{opendata.bristol.gov.uk/datasets/bcc::youth-offenders-in-bristol-by-ward/explore}}: This data records the number and distribution of youth offenders in Bristol by Ward from 2016/17 to 2022/23. 
\end{itemize}

In this study, we perform the following processing steps on the original dataset to construct an analysis dataset for the experiments: First, we extract relevant data from 2016 to 2022 from the above independent data repositories  and integrate them based on the Wards. By matching different aspects of the data in the same Ward in the same year, we ensure the consistency and comparability of the data. Then, we clean the integrated data and finally obtain 2,481 usable data instances.

For categorical features, we use the One-Hot Encoding method to convert them into numerical forms suitable for machine learning models. At the same time, the numerical features are normalized by a Standard Scaler (z-score normalization) to ensure the scale consistency of different features in the model.

\section{Appendix D: The Blind Spots in Fairness Analysis: Intersectional Fairness Perspective}

\begin{figure*}[t]
    \centering
    \includegraphics[width=\linewidth]{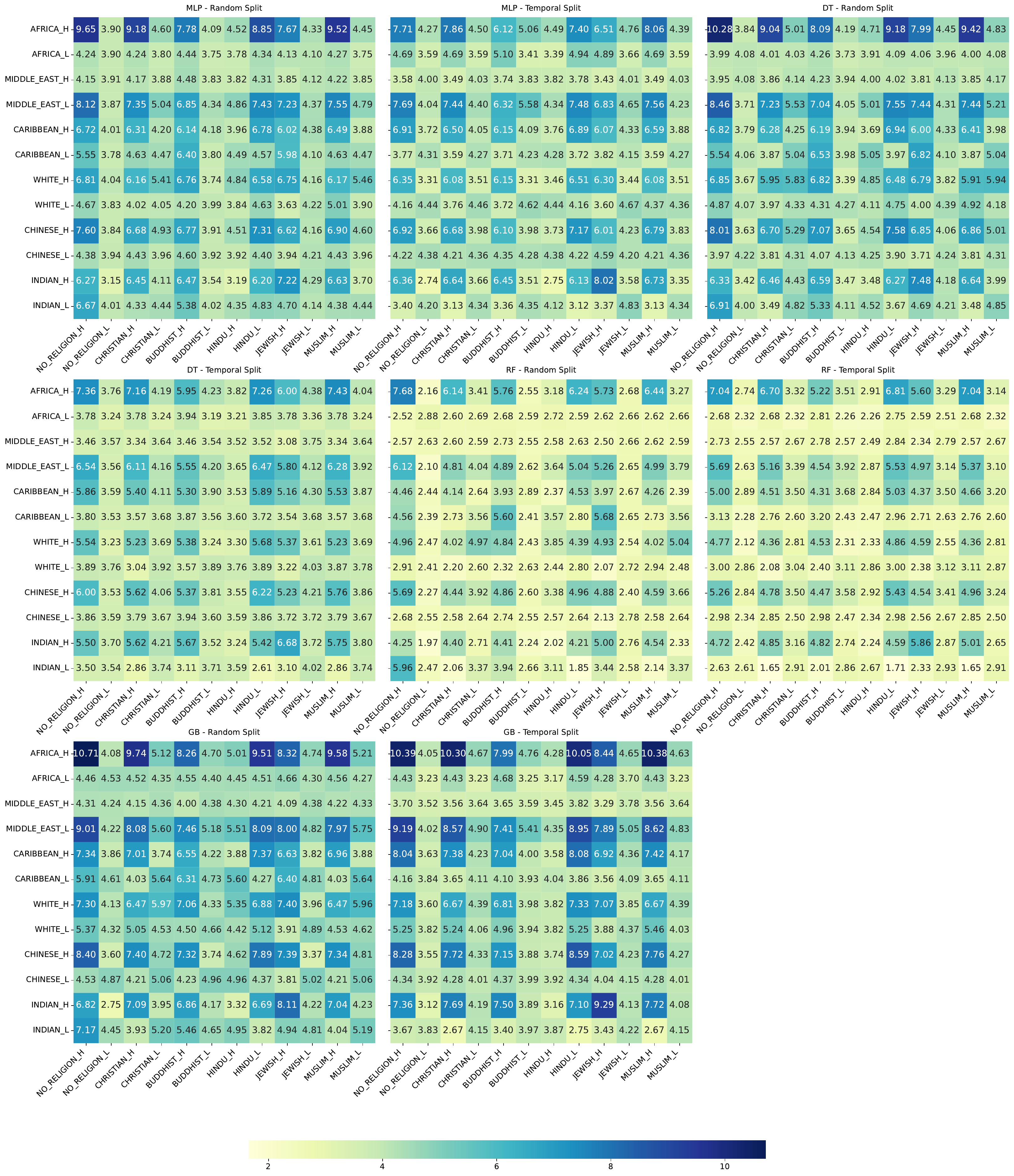}
    \caption{Intersection Feature Fairness Analysis Experiment.}
    \label{fig:intersection}
\end{figure*}

To explore further how the complex relationships between sensitive features can affect unfairness variations, we perform intersectional fairness analysis. We conduct experiments using six features from \(S_{\text{race}}\) and six from \(S_{\text{religion}}\). In this set of experiments, we analyze unfairness not only on a single sensitive feature but also on combinations of race and religion features. This is because different combinations of features may lead to unique patterns of unfairness in the model. For example: the MLP model performs fairly for `\textit{Africa}' in general, but is significantly unfair on the subgroup of `\textit{Africa Muslim}'.  There are 8 experiments on two split methods and four models, the test results are shown in Figure~\ref{fig:intersection}. To better compare the unfairness in the results, we fix the race features and calculate the $\Delta MAE$ of the religion features.

Let $C = \{\textit{Race}, \textit{Religion}\}$ be two distinct sensitive feature classes. For any given feature A1 from \textit{Race} (e.g., Indian) and A2 from \textit{Religion} (e.g., No religion), we measure the intersectional bias as follows:
First, we use formulas~(\ref{eq:A2influence}) to quantify the influence of A2 on prediction bias when A1 is fixed at different level (e.g., Low, High). We then calculate these $\Delta$MAE values for all possible values of A2 (e.g., Muslim, Christian) and take their weighted average as the overall intersectional fairness metric for A1.
An example comparison is shown in Table~\ref{tab:comparison}.

\begin{equation}
\begin{aligned}
\Delta \text{MAE}_{\text{Low[A1]}, A2} &=
\left| 
  \text{MAE}\left(
    D_{\text{Low[A1],} \, 
    \text{High[A2]}}
  \right) \right. \\
&\quad \left.
  -\,
  \text{MAE}\left(
    D_{\text{Low[A1],} \, 
    \text{Low[A2]}}
  \right)
\right| \\
\Delta \text{MAE}_{\text{High[A1]}, A2} &=
\left| 
  \text{MAE}\left(
    D_{\text{High[A1],} \,
    \text{High[A2]}}
  \right) \right. \\
&\quad \left.
  -\,
  \text{MAE}\left(
    D_{\text{High[A1],} \,
    \text{Low[A2]}}
  \right)
\right|
\end{aligned}
\label{eq:A2influence}
\end{equation}

The result indicates a matter of significant concern. While some results align with previous single-feature fairness analysis findings which show consistently low error rates for \textit{White} populations and high bias against minority groups,  deeper investigation uncovers underlying discrimination.
Previous analysis focusing on single-attribute fairness metrics showed remarkably low $\Delta$MAE for \textit{Africa} populations, suggesting minimal algorithmic bias. However, when examined through an intersectional lens, \textit{Africa} individuals experience the highest mean error rates among all intersectional subgroups, indicating severe unfairness that single-feature analysis failed to capture. We are not just missing part of the picture — we are potentially reinforcing discriminatory practices by declaring systems `fair' based on incomplete evaluation criteria.

The prevalent practice of evaluating protected features in isolation creates dangerous blind spots: models that appear fair when evaluated along a single dimension may contain significant biases. By relying solely on traditional single-feature fairness metrics, we risk implementing bias mitigation strategies that leave vulnerable intersectional populations behind. This oversight does not just represent a technical shortcoming—it perpetuates systemic oppression against precisely those groups most in need of protection.

\begin{table*}[t]
\centering

\begin{tabular}{c|cc|cc|cc|cc|cc|cc}

\hline
  & \multicolumn{2}{c|}{India} & \multicolumn{2}{c|}{Chinese} & \multicolumn{2}{c|}{White} & \multicolumn{2}{c|}{Middle East} & \multicolumn{2}{c|}{Caribbean} & \multicolumn{2}{c}{Africa} \\ 
 & High & Low & High & Low & High & Low & High & Low & High & Low & High & Low \\
\hline
No   Religion & 3.12 & 2.66 & 3.76 & 0.44 & 2.77 & 0.84 & 0.24 & 4.25 & 2.71 & 1.77 & 5.75 & 0.34 \\
Christian     & 2.34 & 0.11 & 1.75 & 0.47 & 0.75 & 0.03 & 0.29 & 2.31 & 2.11 & 0.16 & 4.58 & 0.44 \\
Buddhist      & 2.93 & 1.36 & 2.86 & 0.68 & 3.02 & 0.21 & 0.65 & 2.51 & 1.96 & 2.60 & 3.69 & 0.69 \\
Hindu         & 3.01 & 0.48 & 2.80 & 0.48 & 1.74 & 0.79 & 0.49 & 2.57 & 2.82 & 0.08 & 4.33 & 0.56 \\
Jewish        & 2.93 & 0.56 & 2.46 & 0.27 & 2.59 & 0.59 & 0.27 & 2.86 & 1.64 & 1.88 & 3.34 & 0.03 \\
Muslim        & 2.93 & 0.06 & 2.30 & 0.47 & 0.71 & 1.11 & 0.37 & 2.76 & 2.61 & 0.16 & 5.07 & 0.52 \\
\hline
Avg.          & 2.88 & 0.87 & 2.66 & 0.47 & 1.93 & 0.60 & 0.39 & 2.88 & 2.31 & 1.11 & 4.46 & 0.43 \\
$|\Delta \text{Avg.}|$ & \multicolumn{2}{c|}{2.01} & \multicolumn{2}{c|}{2.19} & \multicolumn{2}{c|}{1.34} & \multicolumn{2}{c|}{2.49} & \multicolumn{2}{c|}{1.20} & \multicolumn{2}{c}{4.03}\\
\hline
\end{tabular}
\caption{Intersectional Fairness Performance Comparison under MLP with Random Split. This comparison corresponds to the first subfigure in Figure \ref{fig:intersection}.
The first value 3.12 in the row No Religion is the $\Delta MAE$ which is the difference between 6.27 (INDIAN\_H, NO\_RELIGION\_H) and 3.15 (INDIAN\_H, NO\_RELIGION\_L), and the second value 2.66 is the difference between 6.67 (INDIAN\_L, NO\_RELIGION\_H) 4.01 (INDIAN\_L, NO\_RELIGION\_L).}
\label{tab:comparison}
\end{table*}

\section{Appendix E: Data Fairness Analysis}

The data analysis focuses on the influence of sensitive features, such as race and religion proportions, on the crime rate. For example, the relationship between a high proportion of a specific race and an elevated crime rate is examined. These observations provided insights into the potential presence of systematic unfairness within the data.

\begin{figure*}[t]
    \centering
    \includegraphics[width=\textwidth]{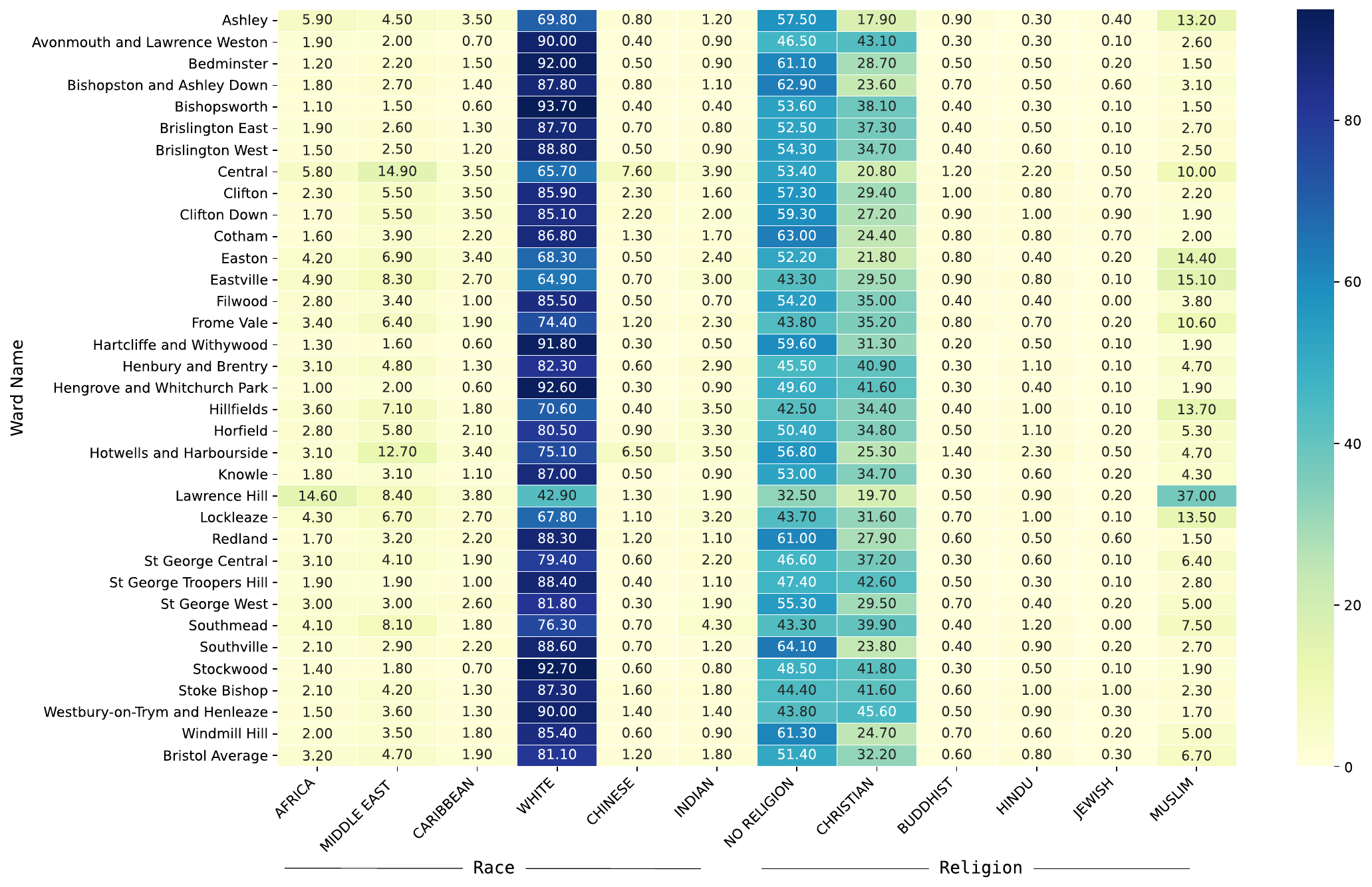}
    \caption{Raw Proportion of Race/Religion Across Wards. The last row `Bristol Average' shows the average proportion across the whole Bristol area. The figure shows the 12 sensitive features selected, which are not representative of all diversities of the possible races and religions in Bristol.}
    \label{fig:race_stat}
\end{figure*}

We select six features from \(S_{\text{race}}\) and six from \(S_{\text{religion}}\), respectively, and the statistics are shown in Figure~\ref{fig:race_stat}. As can be seen, some race and religion groups are not evenly distributed, with extremely high or low proportions in some Wards. For example, the proportion of the \textit{Chinese} in most Wards is less than 1\% with a 1.2\% average proportion in Bristol, but \textbf{Central} has an extremely high value of 7.6\%. Also, most proportions of \textit{White} are between 65\% and 95\%, but it is as low as 42.9\% in the \textbf{Lawrence Hill}.

\begin{figure*}[t]
    \centering
    \includegraphics[width=\textwidth]{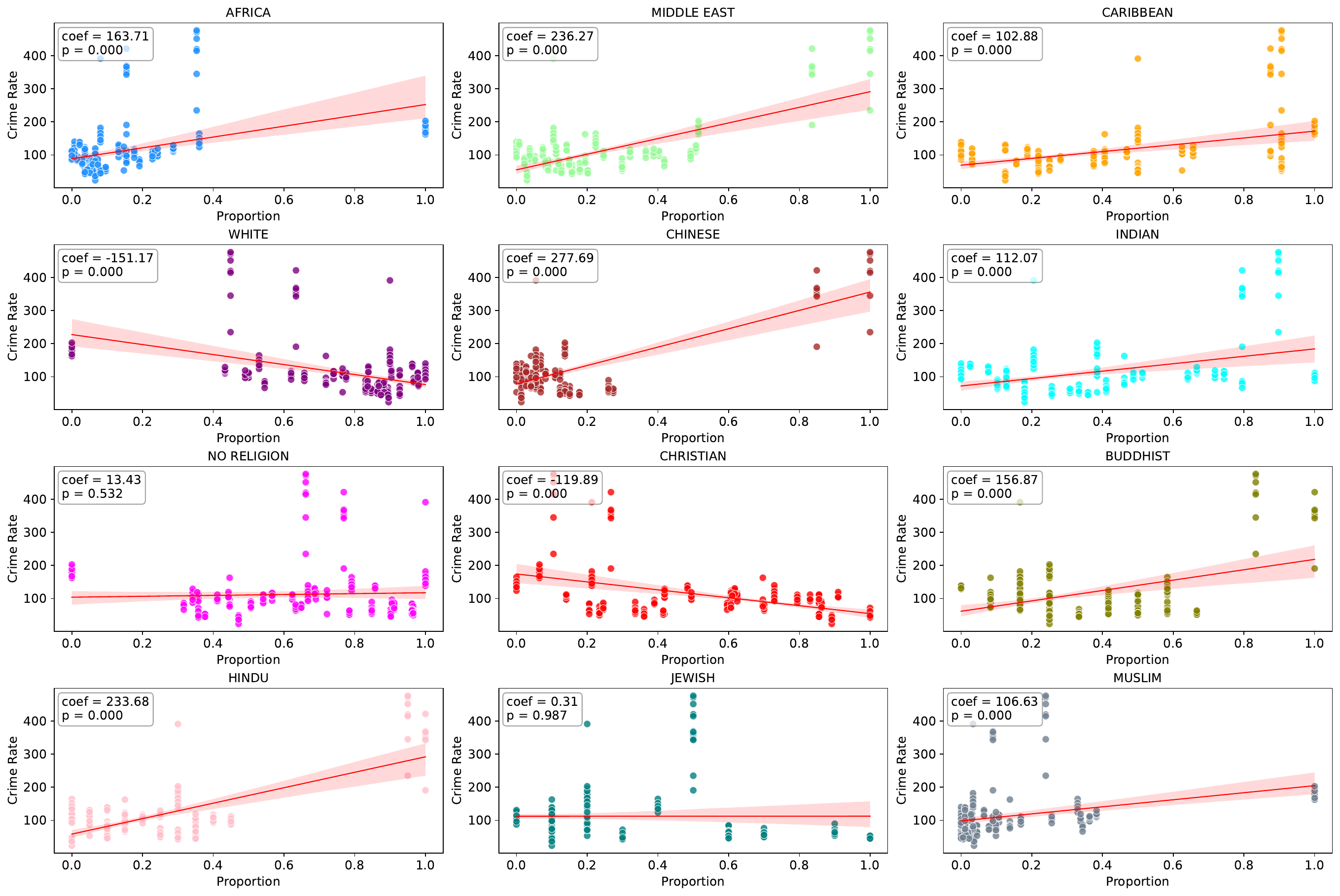}
    \caption{
    Scatter Plot between Sensitive Feature and Crime Rate. A min-max scaler is used to scale the proportion of feature value range into [0,1]. A red regression line with a 95\% confidence interval is drawn using seaborn~\cite{waskom2021seaborn} to see if there is a significant linear trend between a certain race/religion proportion and crime rate (the number of crimes per 1,000 people), with displaying the regression coefficient and p-value .}
    \label{fig:race}
\end{figure*}

The distribution of sensitive features is visualized in a scatter plot (Figure~\ref{fig:race}), alongside the crime rates (all types) across each Ward. A min-max scaler is applied to scale the ratios of each race and religion into the same proportions range [0, 1], as the original proportions of Black and minority ethnic and other-religion are significantly lower compared to those of \textit{White}, \textit{Christian}, and proportions with \textit{No Religion}. From Figure~\ref{fig:race}, we can observe the Wards with a higher proportion of minority ethnic tend to have higher crime rates, while those with a higher proportion of \textit{White} exhibit lower crime rates. In terms of religion, the proportions of \textit{No Religion} and \textit{Jewish} appear to have no significant correlation with crime rates. However, for \textit{Buddhist}, \textit{Hindu}, and \textit{Muslim}, crime rates show a positive correlation with their proportions, whereas \textit{Christian} demonstrates a negative correlation with crime rates.

Although the previous report shows that ethnic segregation in England has reached an all-time low when measured by the number of different groups~\cite{Nagesh2023EthnicSegregation}, the issue of distributional imbalance remains quite pronounced when considering absolute population quantity.  Investigating the impact of uneven distribution becomes important.
The unevenly distributed proportions of sensitive features on different Wards and the difference between sensitive features and crime rates can only indicate whether there is a potential bias in the data. However, it is just a part of the fairness analysis, when building and evaluating the models, it is still necessary to pay attention to differences in model performances in different sensitive features. Therefore, in the following, we focus on the fairness evaluation of model performance and the potential for unfair decision-making.

\end{document}